\documentclass[fleqn,12pt,twoside]{article}
\usepackage{espcrc1}

\usepackage{graphicx}
\usepackage[figuresright]{rotating}

\newcommand{\scp}{\bf P } 
\newcommand{\scc}{\cal C} 
\newcommand{\qbar}{\bar{q}}

\newcommand{\Lam}{\Lambda} 
\newcommand{\lam}{\lambda}

\newcommand{\Ht}{\lambda_{\overline{2}}\lambda_{4};\lambda_{1}\lambda_{\overline{3}}} 

\title{A charge-conjugation-invariance constrained Pomeron-quark coupling}  

\author{Lon-chang Liu
\address{Theoretical Division, Los Alamos National Laboratory, \\  
        Los Alamos, NM 87545, U.S.A.} }%
       
\begin{document}

\maketitle

\begin{abstract}
The commonly used   
$\gamma_{\mu}$ Pomeron-quark coupling changes its sign   
under charge conjugation, in contradiction to the property of Pomeron.        
I show that the Pomeron-quark coupling is tensorial and is invariant     
under the charge conjugation.   
\end{abstract}

\section{Introduction}

High-energy diffractive processes have been extensively modeled  
with Pomeron($\scp$) exchange. It was   
shown \cite{Land} that the $\scp$-proton coupling to a very good approximation
is given by the sum of individual $\scp$-quark coupling and that  
in the two-gluon model of the Pomeron 
the $\scp$-quark coupling is  
proportional to $\gamma_{\mu}$, similar to that of  
C=1 isoscalar photon.   
As noted in Ref.~\cite{Pich}, the model has a dubious aspect because    
$\gamma_{\mu}$ is odd under charge conjugation, ${\scc}$, while 
the Pomeron should be even under $\scc$. In spite of this, 
the $\gamma_{\mu}$-model has been extensively employed 
~\cite{Pich}-\cite{Lage}.  
 
In this work, I use  
the nonperturbative nature of the multigluon exchange and treat the Pomeron 
as a bound state of the gluons. As a result, a tensorial   
${\scp}$-quark coupling emerges, which has the correct   
charge-conjugation property. My study also shed light on 
why the $\gamma_{\mu}$ coupling can be successful inspite of 
its intrinsic inadequacy.    

\section{Doorway-state model for ${\scp} q\qbar$ coupling}  

The doorway model for the $s$-channel $qq$ scattering and the corresponding    
$t$-channel $q\qbar$ scattering is illustrated in Fig.1, where    
the doorway-state (shaded rectangle) is required to 
possess the quantum numbers of a Pomeron.
At small $t$'s the  
Pomeron trajectory is linear:  
$ \alpha(t) = \alpha_0 + \alpha' \ t \ $.   
The spins of the Pomerons are given by 
$J ={\cal R}e [\alpha(t)] = \alpha_0 + {\cal R}e[\alpha']\ t \ ,$   
where  $\alpha_0=1.08$ and ${\cal R}e[\alpha']=0.20\pm 0.02$  
~\cite{Coll},\cite{Bloc}, and  0.25~\cite{Donn}.  
The spins compatible with the  
general properties of the Pomeron   
are $J=2, 4, 6, ...$.  
For definiteness, I will discuss the case with $J=2$.   
The inclusion of higher $J$ in the theory is straightforward. 

It suffices to analyze the lower ${\scp}q\qbar$ vertex in Fig.1(b), where 
$p^{(t)}_2= -p'_1$ and $p'^{(t)}_1 = -p_2$ are the momenta of the $\qbar$. 
The LSZ reduction gives the vertex function  
\begin{equation} 
X^{(t)} \propto   
\bar{u}(p'_2,\ s')\ \Omega^{(t)}_{\rho\sigma}\ v(p'^{(t)}_1,\ s^{(t)})\ e^{\rho\sigma}(P,\Lambda)  ,      
\label{eq:2.4} 
\end{equation} 
where $e^{\rho\sigma}$ is the spin-2 tensor spinor. 

\vspace{-0.5cm} 
\noindent
\begin{figure}[htb] 
\begin{center} 
\includegraphics[height=6.0cm]{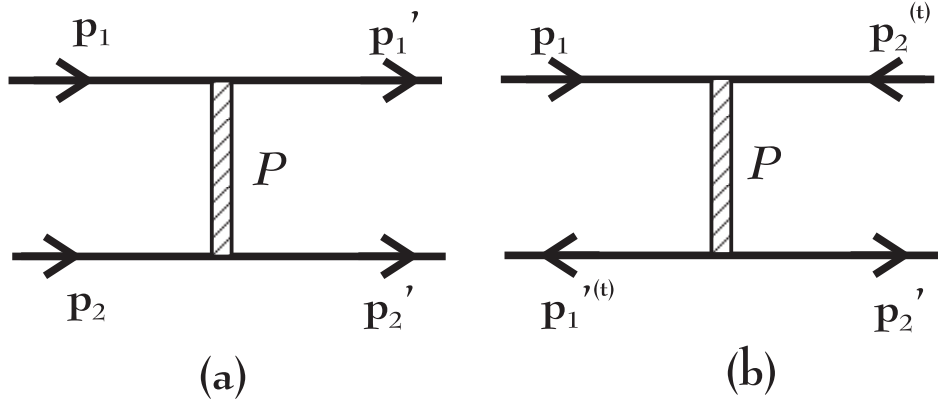} 
\vspace{-1.0cm} 
\caption{Doorway model. The solid lines and the shaded bar denote, respectively,
the $q$ or $\qbar$ and the Pomeron: 
(a) $s$-channel $qq$ scattering;  
(b) $t$-channel $q\qbar$ scattering.} 
\end{center}
\vspace{-0.5cm} 
\end{figure} 

Since $e^{\rho\sigma}$ is a symmetric tensor and $\bar{u}v$ is scalar, it 
follows that $\Omega^{(t)}_{\rho\sigma}$ is a symmetric tensor.  
The most general expansion of $\Omega^{(t)}_{\rho\sigma}$ is 
\begin{equation} 
\Omega^{(t)}_{\rho\sigma} = a_{(\rho)}\ g_{\rho\sigma} + h_{\rho\sigma} + (b_{\rho}\gamma_{\sigma} + b_{\sigma}\gamma_{\rho})  
 +  c\ \sigma_{\rho\sigma} + (d_{\rho}\gamma_{\sigma} + d_{\sigma}\gamma_{\rho})\gamma_5 + f_{\rho\sigma}\gamma_5     
\label{2.6} 
\end{equation} 
where $a_{(\rho)}$ and  $c$ are scalars, $h_{\rho\sigma}$ a
symmetric tensor, $b_{\rho}$ and $b_{\sigma}$ 4-vectors, $d_{\rho}$ and 
$d_{\sigma}$ 
pseudo 4-vectors, and $f_{\rho\sigma}$ a pseudo tensor. 
Only two of the three 4-vectors in Eq.(\ref{eq:2.4}) are 
independent. Because it is not possible to construct either a pseudoscalar or a 
pseudovector, or a pseudotensor from two 4-vectors, it follows that 
$d_{\rho}=d_{\sigma}=0$ and $f_{\rho\sigma}=0.$\ Furthermore, since  
the Pomeron has no charge and since $\bar{u}v$ is 
invariant under $\scc$, it follows that $\Omega^{(t)}_{\rho\sigma}$ is 
also invaraint with respect to $\scc$. Because $\gamma_{\rho}, 
\sigma_{\rho\sigma}$ are both odd under $\scc$, hence 
$b_{\rho}=b_{\sigma}=0$ and $c=0$. The final result is 
\begin{equation} 
\Omega^{(t)}_{\rho\sigma} = a_{(\rho)}\ g_{\rho\sigma} + h_{\rho\sigma} \ . 
\label{eq:2.7} 
\end{equation} 
One can write  
$h_{\rho\sigma} = p'^{(t)}_{1\rho}p'_{2\sigma} + p'_{2\rho}p'^{(t)}_{1\sigma}.
$
In the c.m. of the $t$-channel, $p'^{(t)}_{1} = (E(\kappa), -\vec{\kappa})$ 
and $p'_{2} = (E(\kappa), +\vec{\kappa})$. Thus, 
$h_{00}= 2 E^2$, 
$h_{0j}=h_{j0} = 0,$ 
$h_{ij}=h_{ji} = -2\kappa_i\kappa_j\ .$ 

The absence of $\gamma_{\mu}$ in Eq.(\ref{eq:2.7}) is due to the spin-2 tensor 
spinor of the Pomeron which makes $\Omega^{(t)}_{\rho\sigma}$ invariant   
under the $\scc$ operation. 
One should note that in Ref.\cite{Land} no definite spin and parity    
were projected out from the two-gluon state. 
As a result, the two-gluon model led to
a $\gamma_{\mu}$ coupling.
In other words, it is the spin-parity of the Pomeron  
that constrains the symmetry property of the vertex. Of course,  
obtaining a bound state starting with multigluons is still an 
unsolved nonperturbative dynamics. Our doorway approach  
represents an alternative solution.

\section{The $q\qbar$ amplitude} 

The amplitude is equal to         
$\sum_{\rho\sigma\mu\nu} \Omega^{(t)}_{\rho\sigma}\ \Pi^{\rho\sigma,\mu\nu}(P)\  
 \Omega^{(t)^{\dagger}}_{\mu\nu}\equiv {\cal I}_{q\qbar},   
$ 
where $\Pi^{\rho\sigma,\mu\nu}$ is the spin-2 projector.
In the rest frame of the $\scp$, $P=(M, \vec{0})$. 
Hence, 
$ \Pi^{\rho\sigma,\mu\nu}(P)$ reduces to 
$\Pi^{ij,mn} = (\delta_{im}\delta_{jn} + \delta_{jm}\delta_{in})/2 -\delta_{ij}\delta_{mn}/3 \ \ (i,j,m,n=1,2,3).$   
If $a_{(\rho)}= a$, then   
\begin{equation}
{\cal I}_{q\qbar} = \sum_{ijmn} \Omega^{(t)}_{ij} \Pi^{ij,mn} \Omega^{(t)^{\dagger}}_{mn} 
= \frac{8}{3}\mid \vec{\kappa}\mid^{4} \equiv G^2 \ .        
\label{eq:2.14}
\end{equation} 

The effective interaction Lagrangian density of the doorway model is
\begin{equation} 
{\cal L}_{I} = \frac{f}{M} \theta_{\mu\nu}\Phi^{\mu\nu} + g \overline{\psi}\gamma_{\mu}\psi\ A_{\mu} + h.c.     
\label{eq:2.17} 
\end{equation} 
where $M$ and $\Phi^{\mu\nu}$ are the mass and the tensor field of 
the $2^{++}$ Pomeron, $\psi$ the quark field, $A_{\mu}$ the gluon
field, and $f, g$ the coupling 
constants. The color index is omitted but understood. 
The gauge invariant gluonic current is \cite{Nari}~   
$ \theta_{\rho\sigma} = -G^{\epsilon}_{\rho}G_{\epsilon\sigma} 
+\frac{1}{4}g_{\rho\sigma}G_{\delta\xi}G^{\delta\xi} \ .$   
In the abelian approximation 
$G_{\epsilon\sigma} = \partial_{\epsilon}A_{\sigma} -\partial_{\sigma}A_{\epsilon}$ 
and $G^{\epsilon}_{\rho}= \partial_{\rho}A^{\epsilon} - \partial^{\epsilon}A_{\rho}.$  
The leading-order  
diagrams are shown in Fig.2. Detailed evaluation of these diagrams 
supports the ansatz $a_{(\rho)}=a $ 
that led to Eq.(\ref{eq:2.14}).

\vspace{-0.5cm} 
\noindent
\begin{figure}[htb] 
\begin{center} 
\includegraphics[height=6.0cm]{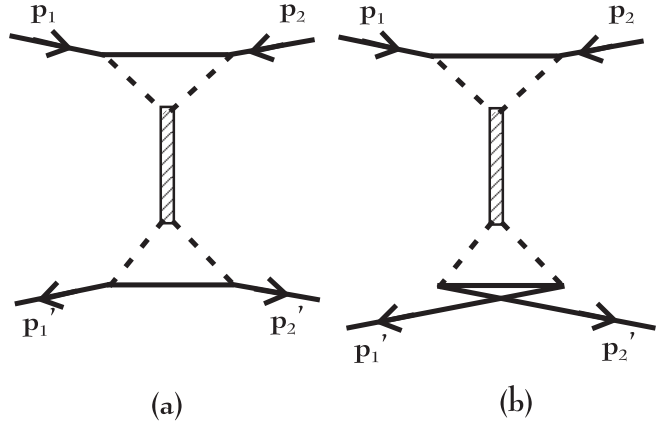} 
\vspace{-1.0cm} 
\caption{Leading-order diagrams of 
$q_{p_1}\qbar_{p_2}\rightarrow q_{p'_1}\qbar_{p'_2}$. The 
dashed lines denote the intermediate gluons.  
There are two additional diagrams (not shown) corresponding,
respectively, to having the    
$q$ and $\qbar$ lines crossed in the upper vertex of (a) and (b). } 
\end{center} 
\vspace{-0.5cm} 
\end{figure} 

The determination of the tensor-coupling vertex  
from the $pp$ scattering data by means of Regge theory can be found in 
Ref.\cite{Liu1} where for the first time   
a singularity-free, 
crossing symmetric, invariant form factor has been formulated and applied
with success.  
The results obtained from fitting $pp$ total and differential cross sections 
at $\sqrt{s}$ between 30 and 60 GeV/c with the use of  
$\alpha'=0.20; J=2$ are: $\Lam_s$=0.65-0.66; $\Lam_t$=1.95-1.96; 
$g_1=1.03-1.04$; $g_3=3.48-3.52$ (all in GeV/c\ ).       
(The dimensions and values of $g_1$ and $g_3$   
differ from those in \cite{Liu1} because of using     
a new parametrization of the amplitude, namely,  (a)    
$M_{new} \equiv (k_s/\sqrt{s}) M_{\Ht}$ and (b) 
in $M_{\Ht}$ the factor $4m^2$ is absorbed into $G_{\lam}G_{\lam'}.$
See eqs.(1) and (2) of \cite{Liu1}-1 for the notation.)  
The form factor is illustrated in Fig.3.

\section{Conclusion} 

The spin and parity of the Pomeron leads to  
a $\scc$-invariant tensor-coupling ${\scp}q\qbar$ vertex  
and to a $q\qbar$ amplitude proportional to   
$G^2$, a scalar.    
One notes that the $\gamma_{\mu}$ 
coupling gives an amplitude $\propto  
 \gamma_{\mu}\otimes\gamma^{\mu}$ which also gives   
a scalar number\cite{Land}.   
This could be the reason that the $\gamma_{\mu}$ model can  
describe the $pp$ scattering data inspite of its inconsistency with the
Pomeron property. 
The tensor coupling should be used because of its 
transformation property under the charge conjugation.   
Study of diffractive processes involving more than two Pomeron vertices 
would differentiate the predictive powers of the $\gamma_{\mu}$ and tensor 
couplings. 

\begin{figure}[t] 
\begin{center}  
\includegraphics[height=9.5cm]{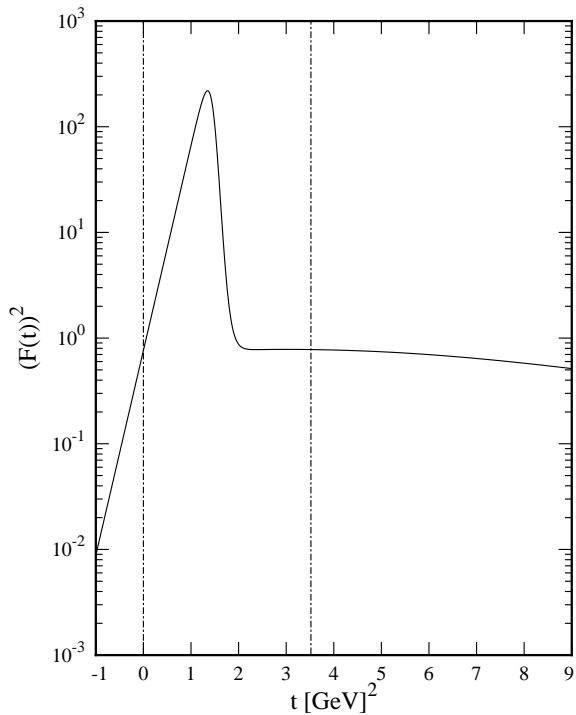} 
\caption{The $t$-dependence of the 
square of the angular-momentum-independent part of the 
singularity-free and crossing-symmetric form factor.  
The physical domains of the $s$- and $t$-channels correspond to 
$t\leq 0$ and $t\geq 4m^2_p=3.52 $[GeV/c]$^2$, respectively.  } 
\end{center} 
\vspace{-0.5cm} 
\end{figure}

\end{document}